\documentclass[aps,twocolumn,
showpacs,preprintnumbers,amsmath,amssymb,pre]{revtex4-1}
\usepackage{epsf,amsmath,amssymb,amsfonts,verbatim,color,pifont}
\usepackage{bm}
\usepackage{graphicx}

\begin{document}
\title{Information thermodynamics for a multi-feedback process with time delay}
\author{Chulan Kwon$^{1,2}$} 
\author{Jaegon Um$^{2}$}
\author{Hyunggyu Park$^{2,3}$}
\affiliation{$^{1}$Department of Physics, Myongji University, Yongin,
Gyeonggi-Do 17058, Korea}
\affiliation{$^{2}$Quantum Universe Center, Korea Institute for Advanced Study, Seoul 02455, Korea}
\affiliation{$^{3}$School of Physics, Korea Institute for Advanced Study, Seoul 02455, Korea}

\date{\today}

\begin{abstract}

We investigate a measurement-feedback process of repeated operations with time delay.
During a finite-time interval, measurement on the system is performed and  the
feedback protocol derived from the measurement outcome is applied with time delay. This
protocol is maintained into the next interval until a new protocol from the next measurement
is applied.
Unlike a feedback process without delay, both memories associated with previous and present measurement outcomes
are involved in the system dynamics, which naturally brings forth
a joint system described by a system state and two memory states. The thermodynamic second law provides a lower bound
for heat flow into a thermal reservoir by the (3-state) Shannon entropy change of the joint system.
However, as the feedback protocol depends on memory states sequentially, we can deduce a tighter bound for heat flow by integrating out
irrelevant memory states during dynamics.
As a simple example, we consider the so-called cold damping feedback process where the velocity of a particle is measured and
a dissipative feedback protocol is applied to decelerate the particle. We confirm that the heat flow is well above
the tightest bound.
We also examine the long-time limit of this feedback process, which turns out to exhibit an interesting instability transition
as well as heating by controlling parameters such as measurement errors, time interval, protocol strength, and time delay length.
We discuss the underlying mechanism for instability and heating, which might be unavoidable in reality.\end{abstract}

\maketitle

The recent information thermodynamics has been proven to resolve the paradox of Maxwell's demon~\cite{maxwell} which was a long-lived problem in spite of enormous research
works~\cite{maxwell,szilard,brillouin,landauer,leff_rex}. Replacing Maxwell's demon by a physical memory device, that was refined by Landauer~\cite{landauer}, one is able to describe
measurement inside a memory device and feedback after measurement acting on the system (engine) as thermodynamic processes. In the measurement process, information acquisition is realized as mutual information gain in the entropy of the joint system  (system and memory device). In the subsequent feedback  process, mutual information is expended through   relaxation out of initial state producing work outside. The work production is balanced energetically by
heat dissipation into the reservoir, which may be negative like in the Szilard engine~\cite{ szilard},
resulting in entropy loss in the reservoir. It was shown that such entropy loss in the reservoir, if any, be compensated sufficiently by the entropy gain of the joint system through mutual information decrease so as to satisfy the second law of thermodynamics. Hence the paradox of Maxwell's demon is resolved. It is the main feature of the information thermodynamics developed by Sagawa and Ueda~\cite{sagawa, sagawa_new, sagawa_network, sagawa_bipartite}. The increase of the total entropy of the joint system and reservoir was proven with the aid of the fluctuation theorem (FT), which was discovered about two decades ago and has been regarded as a principle of nonequilibrium statistical mechanics~\cite{evans,jarzynski,crooks,kurchan,lebowitz}. The role of mutual information in feedback processes has also been confirmed in experiments~\cite{toyabe,koski}.

Memory is usually assumed to reach local equilibrium so fast that system state does not change during measurement. In a feedback process, system state changes in time subject to a fixed protocol given from memory state picked out of its local equilibrium. In this sense, measurement and feedback can reasonably be regarded as processes with separated time periods~\cite{sagawa_bipartite, UHKP} and the fluctuation theorem for the total entropy production was shown to hold separately for the two bipartite periods ~\cite{shiraishi}.

In real situations, however, measurement process takes a finite time and the feedback protocol ought to be applied afterwards. This naturally generates time gap between the start of measurement and feedback. In the present work, we consider a realistic feedback process composed of multiple steps repeated in a finite-time interval, in each of which a feedback protocol is applied with time delay. As an example, we consider a simple cold-damping problem where the velocity of a particle is measured and a dissipative protocol is applied. In repeated feedback steps, the temperature of the system is expected to be cooled down below the reservoir temperature.
\begin{figure}
\includegraphics[width=\columnwidth]{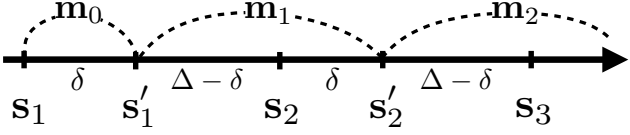}
\caption{(Color online) A schematic picture for repeated measurement-feedback processes. $\mathbf{m}_i$ is a measurement outcome for an initial state $\mathbf{s}_i$ of step $i$, which is applied as a protocol
with time delay $\delta$. This protocol is maintained into the next step until the next protocol is applied.
$\mathbf{m}_0=\mathbf{0}$ when there is no previous measurement.}
\label{fig1}
\end{figure}

Consider that both system state $\mathbf{s}(t)$ and memory state $\mathbf{m}(t)$ in $d$ dimensions coevolve in time $t$ by their own dynamics. At measurement time $t=t_i$, memory starts to measure or copy $\mathbf{s}(t_i)=\mathbf{s}_i $ that acts as a protocol to drive memory into a copied state. One may think of the Langevin dynamics for such a process:
$\dot{\mathbf{m}}=-\tau_{\textrm{m}}^{-1}(\mathbf{m}-\mathbf{s}_i)+\boldsymbol{\xi}(t)$
where $\langle {\xi}_a(t){\xi}_b(t')\rangle=2\tau_{\textrm{m}}^{-1} T_{\textrm{m}}\delta_{ab}\delta(t-t')$ 
with component indices $a,b=1,\cdots, d$ and temperature $T_{\textrm{m}}$ of the reservoir surrounding memory.  The Boltzmann constant is set to unity here and also in the following. Waiting for a long enough time $\delta$ compared to relaxation time $\tau_{\textrm{m}}$, the memory reaches a local equilibrium with the 
conditional probability density function (PDF) given as $p_{\textrm{m}}(\mathbf{m}_i|\mathbf{s}_i)=(2\pi \sigma)^{-d/2} e^{-(\mathbf{m}_i-\mathbf{s}_i)^2/(2\sigma)}$ for $\sigma=\tau_{\textrm{m}} T_{\textrm{m}}$,
which can be interpreted as measurement probability.
 For this period,  the system undergoes a transition to state $\mathbf{s}_i'$ at $t=t_i+\delta$ under a previous protocol $\mathbf{m}_{i-1}$. A new protocol $\mathbf{m}_i$ chosen from the distribution $p_{\textrm{m}}(\mathbf{m}_i|\mathbf{s}_i)$ is applied in turn to the dynamics of the
system for $t_i+\delta <t< t_{i+1}=t_i+\Delta$.  Since the measurement process at step $i$ depends only on $\mathbf{s}_i$,
as seen in the above Langevin equation, intermediate memory states between $\mathbf{m}_{i-1}$ and $\mathbf{m}_i$ can be averaged out for $t_i-(\Delta-\delta) \le t <t_i+\delta$ without influencing the dynamics of $\mathbf{s}(t)$.
In Fig.~\ref{fig1}, the corresponding path of $\mathbf{s}(t)$ is shown with $\mathbf{m}_{i-1}$ and $\mathbf{m}_i$ coexisting in step $i$.

We introduce an adjoint dynamics with time-reverse protocols in which the probability of the system tracing the time-reverse path conjugate to a given (forward) path will be considered.  The time-reversed path is defined as  $\bar{\mathbf{s}}(t)=\varepsilon\mathbf{s}(t_N+t_1-t)$ conjugate to a (forward) path $\mathbf{s}(t)$, where $\varepsilon$ is the parity operator  giving $+1$ ($-1$) if it is applied to a even (odd) parity state in time reversal such as position (momentum). The time-reverse protocols are defined as $\bar{\mathbf{m}}_i=\varepsilon\mathbf{m}_{N-i+1}$. For each of time-reverse protocols, not only the order in time is reversed, but also the parity is multiplied, copying a time-reverse state.

Let $\Pi_{\mathbf{s}_i,\mathbf{s}_i'}^{\mathbf{m}_{i-1}}[\mathbf{s}(t)]$ ($\Pi_{\mathbf{s}_i',\mathbf{s}_{i+1}}^{\mathbf{m}_{i}}[\mathbf{s}(t)]$) be the conditional probability for a partial path from  $\mathbf{s}_i$ ($\mathbf{s}_i'$) to $\mathbf{s}_i'$  ($\mathbf{s}_{i+1}$) under a protocol  $\mathbf{m}_{i-1}$ ($\mathbf{m}_{i}$) for $t_i\le t < t_{i}+\delta$ ($t_i+\delta \le t < t_{i+1}$) in step $i$.  Similarly, we define the conditional path probabilities for time-reverse paths and protocols as $\Pi_{\varepsilon\mathbf{s}_{i+1},\varepsilon\mathbf{s}_i'}^{\varepsilon\mathbf{m}_{i}}[\bar{\mathbf{s}}(t)]$ and $\Pi_{\varepsilon\mathbf{s}_i',\varepsilon\mathbf{s}_{i}}^{\varepsilon\mathbf{m}_{i-1}}[\bar{\mathbf{s}}(t)]$. For usual thermodynamic process without feedback, the change in the total entropy of system and reservoir is known as the log-ratio of the path probabilities of the forward and time-reverse path. Extending to the joint system of system and memory, the corresponding {\em total entropy} change may be written as
\begin{eqnarray}
\sum_{i=1}^N\Delta S_{\textrm{tot},i}&=&\prod_{i=1}^N\ln \left[\frac{\rho_i(\mathbf{s}_i)\rho_i(\mathbf{m}_{i-1}|\mathbf{s}_i)p_{\textrm{m}}(\mathbf{m}_{i}|\mathbf{s}_i)}
{\rho_{i+1}(\mathbf{s}_{i+1})\bar{\rho}(\varepsilon\mathbf{m}_{i},\varepsilon\mathbf{m}_{i-1}|\bar{\mathbf{s}}(t))}\right.
\nonumber\\
&&\left.~~~~~~~~\times \frac{\Pi_{\mathbf{s}_i,\mathbf{s}_i'}^{\mathbf{m}_{i-1}}[\mathbf{s}(t)]\Pi_{\mathbf{s}_i',\mathbf{s}_{i+1}}^{\mathbf{m}_{i}}[\mathbf{s}(t)]}
{\Pi_{\varepsilon\mathbf{s}_{i+1},\varepsilon\mathbf{s}_i'}^{\varepsilon\mathbf{m}_{i}}[\bar{\mathbf{s}}(t)]\Pi_{\varepsilon\mathbf{s}_i',\varepsilon\mathbf{s}_{i}}^{\varepsilon\mathbf{m}_{i-1}}[\bar{\mathbf{s}}(t)]}
\right]
\nonumber\\
&=&\sum_{i=1}^{N}\left[ \Delta S_{\textrm{sm},i}+\Delta S_{\textrm{env},i}\right]
\end{eqnarray}
where $\Delta S_{\textrm{tot},i}$ denotes the contribution from step $i$ and $\rho_i$ is the PDF at $t=t_i$.
A conditional probability $\bar{\rho}$ for time-reverse protocols in the adjoint dynamics can be chosen in  various ways, which will be discussed later. 

The environmental entropy production for step $i$ is defined as
\begin{equation}
\Delta S_{\textrm{env},i}=\ln \left[\frac{\Pi_{\mathbf{s}_i,\mathbf{s}_i'}^{\mathbf{m}_{i-1}}[\mathbf{s}(t)]\Pi_{\mathbf{s}_i',\mathbf{s}_{i+1}}^{\mathbf{m}_{i}}[\mathbf{s}(t)]}
{\Pi_{\varepsilon\mathbf{s}_{i+1},\varepsilon\mathbf{s}_i'}^{\varepsilon\mathbf{m}_{i}}[\bar{\mathbf{s}}(t)]\Pi_{\varepsilon\mathbf{s}_i',\varepsilon\mathbf{s}_{i}}^{\varepsilon\mathbf{m}_{i-1}}[\bar{\mathbf{s}}(t)]}
\right]~.
\label{env_EP}
\end{equation}
In the absence of odd-parity states, $\Delta S_{\textrm{env},i}$ is equal to $Q_i/T$ for heat production $Q_i$ into the reservoir at temperature $T$. However, it may contain an unconventional contribution due to an odd-parity force induced by an odd-parity protocol~\cite{KYKP}.
We will encounter this situation for a cold-damping problem where the velocity of a particle is measured.

$\Delta S_{\textrm{sm},i}$ is the entropy change of the joint system for step $i$, which reads $\Delta S_{\textrm{sys},i}-\Delta I_i$. Here $\Delta I_i$ is the mutual information change between system and memory. Note that the memory state does not change during each step. 
We find the first term to be the Shannon entropy change of the system, given as
\begin{equation}
\Delta S_{\textrm{sys},i}=-[\ln \rho_{i+1}(\mathbf{s}_{i+1})-\ln \rho_i(\mathbf{s}_i)],
\label{shannon}
\end{equation}
resulting from choosing the initial PDF of the time-reverse dynamics  to be the final PDF $\rho_{i+1}(\mathbf{s}_{i+1})$ of the given dynamics. $\Delta I$ depends on how $\bar{\rho}{(\varepsilon\mathbf{m}_i,\varepsilon\mathbf{m}_{i-1}|\bar{\mathbf{s}}(t))}$ is chosen  in the time-reverse dynamics.

We consider two choices in setting the distribution of protocols in the time-reverse dynamics, each of which yields mutual information as a part of $\Delta S_{\textrm{sm},i}$. The first one is given by
\begin{equation}
\bar{\rho}(\varepsilon\mathbf{m}_i,\varepsilon\mathbf{m}_{i-1}|\bar{\mathbf{s}}(t))=\rho_{i+1}(\mathbf{m}_{i-1},\mathbf{m}_i|\mathbf{s}_{i+1})~,
\end{equation}
which is the conditional PDF of the joint system at time $t_{i+1}$ for the given dynamics found as $\rho_{i+1}(\mathbf{s}_{i+1},\mathbf{m}_{i-1},\mathbf{m}_i)\rho_{i+1}(\mathbf{s}_{i+1})^{-1}$. Then, we have
\begin{equation}
\Delta I_i^{(1)}=\ln\frac{\rho_{i+1}(\mathbf{m}_{i-1},\mathbf{m}_i|\mathbf{s}_{i+1})}{\rho_i(\mathbf{m}_{i-1}|\mathbf{s}_i)p_{\textrm{m}}(\mathbf{m}_i|\mathbf{s}_{i})}~,
\label{MI_1}
\end{equation}
which is the change in mutual information between system and two-state memory. The second choice is
\begin{equation}
\bar{\rho}(\varepsilon\mathbf{m}_i,\varepsilon\mathbf{m}_{i-1}|\bar{\mathbf{s}}(t))
=\rho_{i+1}(\mathbf{m}_i|\mathbf{s}_{i+1})
\rho_{i'}(\mathbf{m}_{i-1}|\mathbf{s}_{i}',\mathbf{m}_{i})
\end{equation}
where the first (second) factor determines the distribution of $\epsilon\mathbf{m}_i$ ($\epsilon\mathbf{m}_{i-1}$) for the period $\Delta -\delta$ ($\delta$) of step $N-i$  in the time-reverse dynamics. Then, we have
\begin{equation}
\Delta I^{(2)}_i= \ln \frac{\rho_{i'}(\mathbf{m}_{i-1},\mathbf{m}_i|\mathbf{s}_i')}{\rho_i(\mathbf{m}_{i-1}|\mathbf{s}_i)p_{\textrm{m}}(\mathbf{m}_i|\mathbf{s}_{i})}+ \ln \frac{\rho_{i+1}(\mathbf{m}_i|\mathbf{s}_{i+1})}{\rho_{i'}(\mathbf{m}_i|\mathbf{s}_{i}')}~,
\label{MI_2}
\end{equation}
where $\rho_{i'}$ is the PDF at $t=t_i+\delta$ and $\rho_{i'}(\mathbf{m}_{i-1},\mathbf{m}_i|\mathbf{s}_i')/\rho_{i'}(\mathbf{m}_i|\mathbf{s}_{i}')=\rho_{i'}(\mathbf{m}_{i-1}|\mathbf{s}_{i}',\mathbf{m}_{i})$ is used. The first term is the change in mutual information between system and two-state memory coexisting in the delay period, and the second is that between system and new memory in the remaining period. Writing
$\Delta S_{\textrm{tot},i}^{(1,2)}=\Delta S_{\textrm{sm},i}^{(1,2)}+\Delta S_{\textrm{env},i}$  with $\Delta S_{\textrm{sm},i}^{(1,2)}=\Delta S_{\textrm{sys},i}-\Delta I_i^{(1,2)}$. We can show
both satisfy the FT such that $\langle e^{-\sum_i\Delta S_{\textrm{tot},i}^{(1,2)}}\rangle=1$ and also $\langle e^{-\Delta S_{\textrm{tot},i}^{(1,2)}}\rangle=1$, leading to the inequality $\langle\Delta S_{\textrm{tot},i}^{(1,2)}\rangle  \ge 0$, the generalized thermodynamic second law.

\begin{figure}
\centering
\includegraphics*[width=\columnwidth]{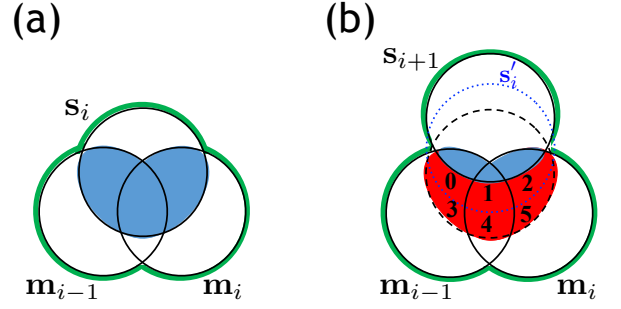}
\caption{(Color online)
Venn diagrams for Shannon entropies (discs) and mutual informations (intersections). $I_i^{(1)}$ is presented by blue areas in (a) at $t=t_i$ and in (b) at $t=t_{i+1}$. The figure (b) presents that initial state $\mathbf{s}_i$ evolves to $\mathbf{s}_i'$ and subsequently to $\mathbf{s}_{i+1}$. $-\Delta I_i^{(1)}$ is represented by the whole red area, $-\Delta I_i^{(2)}$ by the areas labeled by 1, 2, 3, 4, 5, and  $-\Delta I_i^{(3)}$ by those labeled by 1, 2, 3, 4. }
\label{fig2}
\end{figure}

Another choice is given from
\begin{eqnarray}
\Delta S_{\textrm{tot},i}^{\delta(3)}&=&\ln\left[\frac{\rho_i(\mathbf{s}_i)\rho_i(\mathbf{m}_{i-1}|\mathbf{s}_i)
\Pi_{\mathbf{s}_i,\mathbf{s}_i'}^{\mathbf{m}_{i-1}}[\mathbf{s}(t)]}
{\rho_{i'}(\mathbf{s}_i')\rho_{i'}(\mathbf{m}_{i-1}|\mathbf{s}_i')\Pi_{\varepsilon\mathbf{s}_i',\varepsilon\mathbf{s}_i}^{\varepsilon\mathbf{m}_{i-1}}[\bar{\mathbf{s}}(t)]}
\right]~,\\
S_{\textrm{tot},i}^{\Delta-\delta(3)}&=&\ln\left[\frac{\rho_{i'}(\mathbf{s}_i')\rho_{i'}(\mathbf{m}_{i}|\mathbf{s}_i')
\Pi_{\mathbf{s}_i',\mathbf{s}_{i+1}}^{\mathbf{m}_{i}}[\mathbf{s}(t)]}{\rho_{i+1}(\mathbf{s}_{i+1})
\rho_{i+1}(\mathbf{m}_i|\mathbf{s}_{i+1})\Pi_{\varepsilon\mathbf{s}_{i+1},\varepsilon\mathbf{s}_i'}^{\varepsilon\mathbf{m}_{i}}
[\bar{\mathbf{s}}(t)]}\right]~,
\end{eqnarray}
which are defined for $t_i\le t\le t_i+\delta$ and $t_i+\delta\le t\le t_{i+1}$, respectively.
The FT can be shown to hold separately for the two as $\langle e^{\Delta S_{\textrm{tot},i}^{\delta(3)}}\rangle=1$ and $\langle e^{\Delta S_{\textrm{tot},i}^{\Delta-\delta(3)}}\rangle=1$, but not for the sum of them, $\langle e^{-\Delta S_{\textrm{tot},i}^{(3)}}\rangle\neq 1$, for $\Delta S_{\textrm{tot},i}^{(3)}=\Delta S_{\textrm{sm},i}^{(3)}+\Delta S_{\textrm{env},i}$.  However, the inequality holds for the sum, $\langle\Delta S_{\textrm{tot},i}^{(3)}\rangle\ge 0$. We can similarly write $\Delta S_{\textrm{sm},i}^{(3)}=\Delta S_{\textrm{sys},i}-\Delta I_i^{(3)}$ where
\begin{equation}
\Delta I_i^{(3)}= \ln \frac{\rho_{i'}(\mathbf{m}_{i-1}|\mathbf{s}_i')}{\rho_i(\mathbf{m}_{i-1}|\mathbf{s}_i)}+ \ln \frac{\rho_{i+1}(\mathbf{m}_i|\mathbf{s}_{i+1})}{\rho_{i'}(\mathbf{m}_i|\mathbf{s}_{i}')}.
\label{MI_3}
\end{equation}
As presented in Fig.~\ref{fig2}, $-\Delta I_i^{(3)}$ is found to have the lowest bound to the change in total entropy among the three representations. One can say that the total entropy change is overestimated as considered is mutual information between system and protocol having no influence on the dynamics. Overestimated are mutual information due to new protocol in time delay and that due to past protocol in new feedback period, labeled by 0 and 5 in the figure, respectively.

We apply our theory to a cold-damping problem where a feedback force is applied in the opposite direction to the measured velocity~\cite{khkim,jourdan,ito}. From now on, we investigate the problem within a single step, say for $t_1\le t \le t_2$. We consider the one-dimensional motion of a particle described by the Langevin equation for the velocity $v$,
\begin{equation}
\dot{v}=-\gamma v-\tilde{\gamma}y_i +\xi(t)~,
\label{langevin}
\end{equation}
where mass is set to unity. Then, $\mathbf{s}=v$ and $\mathbf{m}_i=y_i$ where $i=0$ ($i=1$) denotes past (new) protocol. $y_0$ is applied for $t_1\le t \le t_1+\delta$ and $y_1$ for the remaining period. This feedback process can be realized in experiment for a colloidal particle with charge $q$ where $\tilde{\gamma}$ is a control parameter for an electric field $E=\tilde{\gamma}y/q$. $\xi$ is a usual stochastic force with mean zero and variance $\langle \xi(t)\xi(t')\rangle=2T\gamma\delta(t-t')$. $\tilde{\gamma}>0$ is used for the purpose of cold damping.

One can find various PDF's and moments recursively given the initial PDF $\rho(v_1, y_0)$ with initial moments,
\begin{equation}
T_1 =\langle v^2_1 \rangle ~,~ P_1 =\langle y_0^2 \rangle~,~
R_1 = \langle v_1 y_0 \rangle~.
\label{initial_mom}
\end{equation}
It is convenient to consider composite states at $t=t_1$ and $t=t_2$, given as $\mathbf{c}_1=(v_1,y_0,y_1)$ and $\mathbf{c}_2=(v_2,y_0,y_1)$. Then,
$\rho(\mathbf{c}_{1})$ is equal to the product of $\rho(v_1, y_0)$ and $p_{\textrm{m}}(y_1|v_1)= (2\pi\sigma)^{-1/2}e^{-(y_1-v_1)^2/(2\sigma)}$. The Onsager-Machlup theory~\cite{onsager} gives the conditional probability for path $v(t)$ from $v(\tau)=v$ to $v(\tau')=v'$ as
$\Pi_{v,v'}^{y_i}[v(t)]\propto
\exp[-(4\gamma T)^{-1}\int_{\tau}^{\tau'}dt\left(\dot{u}+\gamma u\right)^2 ]$, where $u(t)=v(t)+(\tilde{\gamma}/\gamma)y_i$. Then,
the path integral of $\Pi_{v_1,v_1'}^{y_0}[v(t)]\Pi_{v_1',v_2}^{y_1}[v(t)]$ over all paths gives rise to the transition probability of $v(t_2)=v_2$ given a composite state $\mathbf{c}_1$. We find
\begin{equation}
\label{eq:propagator}
\rho(v_2|\mathbf{c}_{1})=\frac{1}{\sqrt{2\pi w_\Delta}}e^{-\left(
v_2- e^{-\gamma \Delta} v_{1} + (\tilde{\gamma}/\gamma) f  \right)^2/(2w_\Delta) },
\end{equation}
where $w_{\Delta}=T(1-e^{-2\gamma\Delta})$ and
\begin{equation}
f = \left(  e^{-\gamma (\Delta-\delta)} - e^{-\gamma \Delta} \right)y_0
+ \left( 1 - e^{-\gamma (\Delta-\delta) } \right) y_1.
\end{equation}
Using this, the PDF of $\mathbf{c}_2$ is given as
\begin{equation}
\rho(\mathbf{c}_2) = \int dv_1 \, \rho(\mathbf{c}_{1}) \rho(v_2|\mathbf{c}_1)
= \sqrt{ \frac{{\rm det} {\mathsf D}_2}{(2\pi)^3}}e^{-\mathbf{c}_2
\mathsf{D}_2 \mathbf{c}_2^{\textrm{t}}/2},
\label{rho_c2}
\end{equation}
where the superscript t denotes the transpose.
Using the property of multi-variate Gaussian integral, the inversion of the matrix $\mathsf{D}_2$ yields six moments such that
\begin{equation}
\label{eq:matrix}
\mathsf{D}_2^{-1}=\left(
\begin{array}{ccc}
\langle v_2^2 \rangle & \langle v_2 y_0 \rangle
& \langle v_2 y_1 \rangle \\
 \langle v_2 y_0 \rangle&  \langle y_0^2 \rangle &  \langle  y_0y_1\rangle\\
 \langle v_2y_1 \rangle &   \langle  y_0y_1 \rangle&   \langle y_1^2 \rangle
\end{array}\right)~,
\end{equation}
which can be found in terms of $T_1$, $P_1$, and $R_1$ given in Eq.~(\ref{initial_mom}).

\begin{figure}
\centering
\includegraphics*[width=\columnwidth]{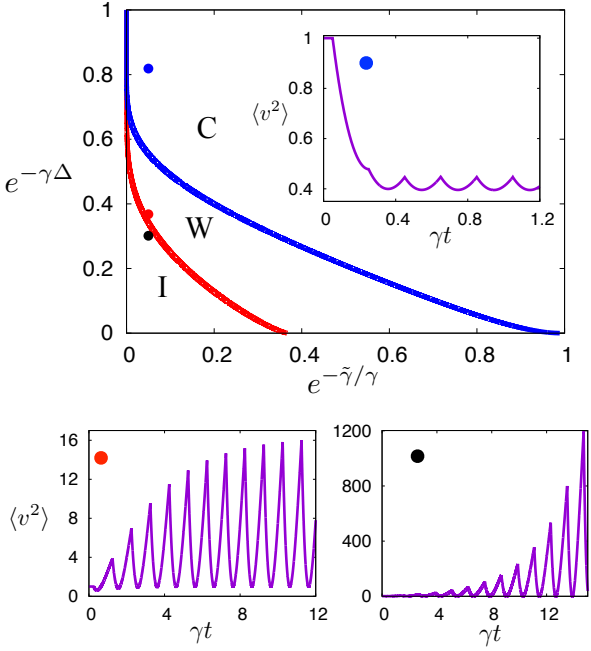}
\caption{(Color online) The diagram is drawn for $\delta / \Delta =0.25$, $\sigma=0.1$. C denotes the region for $T^{\textrm{av}}_{\infty}<T$, W for $T<T^{\textrm{av}}_{\infty}<\infty$, and I for $T^{\textrm{av}}_{\infty}=\infty$. The three points are picked from the three regions, for which $\langle v(t)^2\rangle$ versus $\gamma t$ are shown.}
\label{fig3}
\end{figure}

In particular,  $T_2=\langle v_2^2\rangle$. $P_2=\langle y_1^2\rangle$, and $R_2=\langle v_2 y_1\rangle$ are found to satisfy the linear recursion relation:
\begin{eqnarray}
\label{recursion}
T_2 &=& w_\Delta + \sigma h^2  +K^2 T_1 +L^2 P_1 -2K L R_1, \nonumber\\
P_2 &=& \sigma+T_1, \\
R_2 &=& -\sigma h +K T_1 -L R_1, \nonumber
\end{eqnarray}
where $K = e^{-\gamma \Delta} -H$ with $H=(\tilde{\gamma}/\gamma) \left(1-e^{-\gamma(\Delta-\delta)} \right) $ and $L= (\tilde{\gamma}/\gamma) e^{-\gamma \Delta }\left( e^{\gamma \delta} -1\right) $. The recursion relation can be rewritten as
$\mathbf{Z}_{2}=\mathsf{G}\cdot\mathbf{Z}_1+\mathbf{A}$ for $\mathbf{Z}_i=(T_i,P_i,R_i)^{\textrm{t}}$ where the matrix $\mathsf{G}$ and the vector $\mathbf{A}$ are given from Eq.~(\ref{recursion}). $T_i=\langle v_i^2\rangle$ is defined as the effective temperature at $t=t_i$ and is updated through feedback steps as $T_1\to T_2\to T_3\to\cdots$. The recursion relation will leads to a fixed value $T_{\infty}$ only if $|\lambda_a|<1$ for eigenvalues $\lambda_a$ of $\mathsf{G}$ for $ a=1,2, 3$. The average effective temperature at step $i$ can be found as $T^{\textrm{av}}_i=\Delta ^{-1}\int_{t_i}^{t_{i+1}}dt \langle v(t)^2\rangle$. Cold damping will be successful if $T^{\textrm{av}}_{\infty}<T$.
In Fig.~{\ref{fig3}}, C (cold) stands for the region for $T^{\textrm{av}}_{\infty}<T$, W (warm) for $T^{\textrm{av}}_{\infty}>T$, and I (instability) for the instability region with $|\lambda_a|\ge1$.

\begin{figure}
\centering
\includegraphics*[width=\columnwidth]{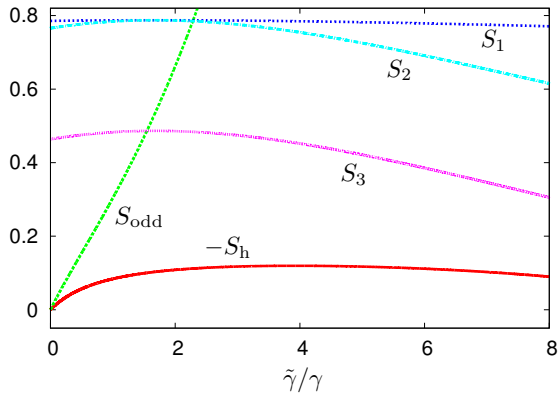}
\caption{(Color online) The components of $\Delta S_{\textrm{tot}}$ as
functions of $\tilde{\gamma}/\gamma$ at the fixed point in the recursion procedure where $T_i=T_{i+1}$. Here, the plot is drawn for $\sigma =0.1$, $\gamma \Delta = 0.2$, $\gamma \delta =0.05$, and $T=1$.
For simplicity, we use $S_i=\langle\Delta S^{(\alpha)}_{\textrm{sm}}\rangle$ for $\alpha=1,2,3$, $S_{\textrm{odd}}=\langle\Delta S_{\textrm{uc}}\rangle$, and $S_{\textrm{h}}=\langle Q/T\rangle$.
}
\label{fig4}
\end{figure}

We can compute the parts of the total entropy change. The Shannon entropy for $\rho(\mathbf{c}_{2})$ in Eq.~(\ref{rho_c2}) can be written as $-\langle \ln \rho(\mathbf{c}_2) \rangle  = (1/2) \left( -\ln {\rm det}\mathsf{D}_2+ 3 \ln 2\pi +3 \right)$, and similarly for $\rho(\mathbf{c}_{1})$. Then, we obtain $\langle \Delta S_{\textrm{sm}}^{(1)}\rangle =\langle\ln [\rho(\mathbf{c}_1)/\rho(\mathbf{c}_2)]\rangle$.
By integrating $\rho(v(t),y_0,y_1)$ over $y_0$ or $y_1$, one can find $\rho(v(t),y_i)$. Then, we find
\begin{equation}
\label{eq:vl}
\left< \ln \frac{\rho(v_1, y_0)}{\rho(v'_1, y_0) }\right> = \frac{1}{2}\ln \left[ e^{-2\gamma \delta }+ \frac{w_\delta P_1}{T_1 P_1 -R_1^2}\right],
\end{equation}
and
\begin{eqnarray}
\label{eq:vn}
\left< \ln \frac{\rho(v_1',y_1)}{\rho(v_2, y_1)}\right> &=&  \frac{1}{2}\ln \left[ e^{-2\gamma (\Delta-\delta) } \right. \\
&&\left. +\frac{w_{\Delta-\delta}(T_1 + \sigma) }{w_\delta T_1
+ \sigma \langle v_{1}'^2 \rangle+ H_{\delta}^2(T_1 P_1 -R^2_1)}\right], \nonumber
\end{eqnarray}
where $H_{\delta} = (\tilde{\gamma}/\gamma) (1-e^{-\gamma \delta} )$. Adding Eqs.~(\ref{eq:vl}) and (\ref{eq:vn}) leads to  $\langle \Delta S_{\textrm{sm}}^{(3)}\rangle $. $\langle \Delta S_{\textrm{sm}}^{(2)}\rangle$ in Eq.(\ref{MI_2}) can be found by adding the one in Eq.~(\ref{eq:vn}) and $\langle\ln \rho(\mathbf{c}_1)- \ln \rho(\mathbf{c}'_1) \rangle$. The three representations of the Shannon entropy change for the joint system are shown in Fig.~\ref{fig4}.

The average environmental entropy production in Eq.~(\ref{env_EP}) is given as
\begin{eqnarray}
\label{eq:se_cold}
\lefteqn{\langle\Delta S_{\textrm{env}}\rangle=
\left\langle \!\! \ln \! \left[\frac{\Pi_{v_1,v_1'}^{y_0}[v(t)]\Pi_{v_1',v_2}^{y_1}[v(t)]}
{\Pi_{-v_2,-v_1'}^{-y_1}[-v(t)]\Pi_{-v_1',-v_1}^{-y_0}[-v(t)]} \!
\right] \! \right\rangle}\\
&=&
\frac{T_{1} -T_{2} }{2T}
-\frac{\tilde{\gamma}}{\gamma T} \bigl [ \left < v_2 y_1 \right>  -\left<  v'_1 y_1 \right> \bigr]
-\frac{\tilde{\gamma}}{\gamma T} \bigl[\left< v'_1 y_0 \right> -\left< v_1 y_0 \right> \bigr ]. \nonumber
\end{eqnarray}
$\langle v'_1 y_1\rangle$, and $\langle v'_1 y_0 \rangle$ can be obtained from $\langle v_2 y_1\rangle$, and $\langle v_2 y_0 \rangle$ in Eq.~(\ref{eq:matrix}) by putting $\Delta=\delta$.

The average heat production is found from $\int_{t_1}^{t_2}dt\langle [\gamma v(t)-\xi(t)]\circ v(t)\rangle$ with $\circ$ denoting the Stratonovich calculus~\cite{Strat}. We find $\langle Q\rangle=\gamma\Delta (T^{\textrm{av}}-T)$. When the average effective temperature is lower than the reservoir temperature, meeting the need of cold damping, the average heat becomes negative, which is the situation in which the paradox of Maxwell's demon is raised.

$\langle\Delta S_{\textrm{uc}}\rangle=\langle\Delta S_{\textrm{env}}\rangle-\langle Q/T\rangle$ is an unconventional entropy production  which is known to appear in the presence of an odd-parity force; $-\tilde{\gamma}y_i$ in our case~\cite{KYKP}. Without feedback control, $\langle \Delta S_{\textrm{tot}}\rangle$ maintains positivity even for a negative $\langle Q\rangle/T$ thanks to $\langle \Delta S_{\textrm{uc}}\rangle$. For feedback process, $-\langle \Delta I\rangle$ plays an additive role in compensating entropy loss in reservoir together with $\langle \Delta S_{\textrm{uc}}\rangle$.

In Fig.~\ref{fig4}, we display the components comprising the total entropy change at the fixed point of the recursive feedback process for $T_1=T_2$ in the above equations. In the figure, $\langle \Delta S^{(\alpha)}_{\textrm{sm}}+\Delta S_{\textrm{uc}}\rangle$ is shown to be greater than $-\langle Q/T\rangle$ for all $\alpha$, which confirms the generalized second law of thermodynamics. As expected from Fig.~\ref{fig2},  $\langle \Delta S^{(3)}_{\textrm{tot}}\rangle$ is shown to yield the tightest bound.

We examine the generalized thermodynamic second law in the presence of coexisting past and present memories. We show the total entropy change to have the tightest bound as only mutual informations influencing the dynamics are considered, which is confirmed in the cold-damping problem.
For the cold-damping using a multi-step feedback, the effective temperature can be reduced below reservoir temperature for a certain range of parameters, while it can reach a higher value or even diverge unlimitedly due to overshooting caused by large $\tilde{\gamma}$ and $\Delta$, as shown in Fig.~\ref{fig3}. We derive the stability condition for the convergence of feedback. An intriguing role of $\delta$ to enhance the stability for large $\Delta$ will be further investigated in a future study~\cite{cold_damping}. We expect overshooting and instability to take place in general feedback processes for finite $\delta$ and $\Delta$, which are unavoidable in reality.

\begin{acknowledgments}
This research was supported by the NRF Grant No.~2013R1A1A2011079 (C.K.) and 2013R1A1A2A10009722 (H.P.).
\end{acknowledgments}

\vfil\eject

\begin{thebibliography}{99}
\bibitem{Strat} For the Wiener process $dW$ defined by $\int_t^{t+\epsilon}ds \xi(s)$ in $\epsilon\to 0$ limit, $\langle dW\circ v(t)\rangle =\langle dW[v(t)+v(t+\epsilon)]/2\rangle=\langle dW\cdot dv/2\rangle$, where $\langle dW\cdot v(t)\rangle=0$ is used. Using $dv\simeq dW$ from Eq.~(\ref{langevin}), $\langle dW\circ v(t)\rangle\simeq \langle (dW)^2\rangle /2=\epsilon\gamma T$.
\end{thebibliography}
\end{document}